\documentclass[%
  twocolumn
]{mpi2015-cscpreprint}

\usepackage{graphicx}      % include this line if your document contains figures
\usepackage{amsmath}
\usepackage{amssymb}

\usepackage{tikz}

\usepackage{pgfplots}
\pgfplotsset{compat = 1.13}

\definecolor{linkBlue}{HTML}{0055C9}
\definecolor{linkRed}{HTML}{FF1A24}
\definecolor{linkPurple}{HTML}{6200D9}

\usepackage{listings}

\usepackage{changepage}

\usepackage{url}

\newtheorem{definition}{Definition}[section]

\newtheorem{remark}{Remark}[section]

\def\IR{{\mathbb R}}
\def\IC{{\mathbb C}}

\def\IL{{\mathbb L}}

\newcommand{\bA}{{\textbf A}}
\newcommand{\bB}{{\textbf B}}
\newcommand{\bC}{{\textbf C}}

\newcommand{\bE}{{\textbf E}}

\newcommand{\bY}{{\textbf Y}}

\newcommand{\bN}{{\textbf N}}
\newcommand{\bI}{{\textbf I}}
\newcommand{\bH}{{\textbf H}}

\newcommand{\bQ}{{\textbf Q}}

\newcommand{\bX}{{\textbf X}}
\newcommand{\bx}{{\textbf x}}

\newcommand{\bu}{{\textbf u}}

\newcommand{\bc}{{\textbf c}}

\newcommand{\bb}{{\textbf b}}

\newcommand{\bv}{{\textbf v}}
\newcommand{\bw}{{\textbf w}}
\newcommand{\bz}{{\textbf z}}

\newcommand{\bff}{{\textbf f}}

\newcommand{\bPhi}{ \boldsymbol{\Phi} }

\newcommand{\cT}{ {\cal T} }

\def\IR{{\mathbb R}}
\def\IC{{\mathbb C}}
\def\IL{{\mathbb L}}
\def\IV{{\mathbb V}}
\def\IW{{\mathbb W}}

\newcommand{\bhA}{{\hat{\textbf A}}}
\newcommand{\bhB}{{\hat{\textbf B}}}
\newcommand{\bhC}{{\hat{\textbf C}}}

\newcommand{\bhE}{{\hat{\textbf E}}}
\newcommand{\bhH}{{\hat{\textbf H}}}
\newcommand{\bhN}{{\hat{\textbf N}}}
\newcommand{\bhQ}{{\hat{\textbf Q}}}

\newcommand{\bhx}{{\hat{\textbf x}}}

\newcommand{\hH}{{\hat{H}}}

\newcommand{\bhPhi}{{\hat{\mathbf \Phi}}}

\newcommand{\chO}{{\hat{\cal O}}}
\newcommand{\chR}{{\hat{\cal R}}}

\begin{document}
\title{A framework for fitting quadratic-bilinear systems with applications to models of electrical circuits}
\novelty{Many classes of nonlinear systems that are described by analytic nonlinearities can be represented equivalently as quadratic-bilinear systems (by means of lifting). The proposed method combines the Loewner framework with the Volterra series theory and constructs reduced quadratic-bilinear systems from input-output time-domain data that approximate the original nonlinear system.}

\author[$\ast$]{Dimitrios S. Karachalios}
\affil[$\ast$]{Max Planck Institute for Dynamics of Complex Technical Systems, Magdeburg, Germany.\authorcr
 \email{karachalios@mpi-magdeburg.mpg.de}, \orcid{0000-0001-9566-0076}}
  
\author[$\dagger$]{Ion Victor Gosea}
\affil[$\dagger$]{Max Planck Institute for Dynamics of Complex Technical Systems, Magdeburg, Germany.\authorcr
  \email{gosea@mpi-magdeburg.mpg.de}, \orcid{0000-0003-3580-4116}}
  
\author[$\ddagger$]{Athanasios C. Antoulas}
\affil[$\ddagger$]{Electrical and Computer Engineering (ECE) Department, Rice University, Houston, USA,\linebreak Max Planck Institute, Magdeburg, Germany, and Baylor College of Medicine, Houston, USA.\authorcr
  \email{aca@rice.edu}}  
  
%\shorttitle{}
%\shortauthor{}
%\date{\today}
%\msc{}
\abstract{In this contribution, we propose a data-driven procedure to fit quadratic-bilinear surrogate models from data. Although the dynamics characterizing the original model are strongly nonlinear, we rely on lifting techniques to embed the original model into a quadratic-bilinear format. Here, data represent generalized transfer function values. This method is an extension of methods that do bilinear, or quadratic inference, separately. It is based on first fitting a linear model with the classical Loewner framework, and then on inferring the best supplementing nonlinear operators, in a least-squares sense. The application scope of this method is given by electrical circuits with nonlinear components (such as diodes). We propose various test cases to illustrate the performance of the method. }

\maketitle  
%%%%%%%%%%%%%%%%%%%%%%%%%%%%%%%%%%%%%%%%%%%%%%%%%%%%%%%%%%%%%%%%%%%%%%%%%%%%%%%%
\section{Introduction}%
\label{sec:intro}
System Identification (SI) and data-driven Model Order Reduction (MOR) are two fairly established methodologies that aim at discovering/computing robust surrogates of dynamical systems from data. This is performed without having an exact access to the system's structure or matrices scaling the various terms. In the case of SI, the aim of discovery of known classes of dynamical systems that constitute an appropriate mathematical formalism capable of describing dynamical phenomena. We refer the reader to \cite{OvMo96} and to \cite{Lju99} for more details on various methodologies. In the case of MOR, the need for approximating the underlying dynamical system is dictated mostly by the increased dimension under hand (the number of internal variables). Conventional MOR methods are indeed intrusive, in the sense that they require an explicit formulation of the dynamical system to be reduced (in terms of matrices or various operators). We refer the reader to \cite{morBenOCetal17} and to \cite{ABG20}, for more details. However, data-driven MOR methods are generally non-intrussive, since they require only data (snapshots of the states, input-output measurements, etc.), and not the full/exact description of the model. Methods that fall into this category range from DMD (dynamical mode decomposition), OpInf (operator inference), to LF (the Loewner framework). Such data-driven methods can be used for identifying surrogate models without having exact access to the original operators. 
%Usually, the underlying evolutionary phenomena can be described formally by a partial differential equation (PDE). Discretization techniques transform the spatial-continuous nature of the PDEs to a finite approximation by employing i.e., finite elements and this result to a system of $n$ ordinary differential equations ODEs. Now, the discretized solution manifold allows infinite approximation to the continuous one after increasing the dimension $n$ reaching easily the curse of dimensionality. As a result, by constructing systems of ODEs with dimension $n\approx 10^9$ makes the computational demands even for the intrusive methods and inside a HPC environment very challenging. 
%Towards introducing non-intrusive methods for dealing with such cases, we want to stress the need for MOR techniques.
Starting with \cite{HOKALMAN}, many algorithms have been developed to identify linear dynamical systems in the state-space realization:
\begin{equation}\label{eq:linsys}
\small
    \left\{\begin{aligned}
        \dot{\bx}(t)&=\bA\bx(t)+\bB u(t),\ \ y(t)=\bC\bx(t),
    \end{aligned}\right.
\end{equation}
where $\bx$ is the state variable of dimension $n$, while the system matrices are given by $\bA\in\IR^{n\times n},~\bB,\bC^T\in\IR^{n\times 1}$. We refer to , for more details on various methodologies. In recent years, the ideas for developing methods for linear systems have steadily expanded to fit certain classes of nonlinear systems (such as polynomial). 
%An important class of such systems is the one that includes polynomial nonlinearities.
We consider nonlinear system can be written in the following state-space representation
\begin{equation}\label{eq:nonlinsys}
\small
\left\{\begin{aligned}  \dot{\bx}(t)&=\bA\bx(t)+\bff(\bx(t),\bu(t))+\bB u(t),\ \ y(t)=\bC\bx(t),
    \end{aligned}\right.     
\end{equation}
where $\bff:\IR^{n} \times \IR \rightarrow  \IR^{n}$ is the nonlinear operator that will be approximated (using Carleman linearization \cite{Car32}) or lifted (McCormick relaxation \cite{Gu11}) to a polynomial structure. In such scenarios, the scaling and superposition principles that hold for linear models, do not hold here anymore, making the analysis fairly difficult. Additionally, many useful phenomena of nonlinear nature can not be accurately approximated by means of linearization methods (that could be performed only locally). Two sub-class of polynomial models that belong to the class described in Eq.\;\eqref{eq:nonlinsys} are the quadratic and the bilinear ones. 
The operators that we will be concerned with in this study are given as follows:
\begin{itemize}
    \item Quadratic case: $\bff_{q}(\bx(t),u(t))=\bQ(\bx(t)\otimes\bx(t))$ and
    \item Bilinear case: $\bff_{b}(\bx(t),u(t))=\bN\bx(t)u(t)$,
\end{itemize}
or with linear combinations of these two classes that appear in the case of quadratic-bilinear (QB) systems. Here, denotes the Kronecker product. For cases when the non-linear operator of the original system is not directly written as $\alpha f_q + \beta f_b$, we can employ lifting techniques to embed the original nonlinear dynamics into the required format (without any approximation whatsoever). This is performed by using specifically tailored lifting
transformations. More specifically, auxiliary variables and equations are introduced in order to reformulate the equations in desired form. This allows to apply conventional MOR methods to more general nonlinear systems. Specific lifting transformations were discussed in \cite{Gu11,morBre13,morKraW19}.

A viable alternative is to employ data-driven methods is the Loewner framework (LF), where the construction of low-order models can be achieved directly from data. It is to be noted that LF has been recently extended to fit certain classes of nonlinear systems from data, such as bilinear systems in \cite{AGI16}, and quadratic-bilinear (QB) systems in \cite{GA18,morAntGH19}. However, in this methods, data used in the computation process can not be easily inferred from practical experiments (it is challenging to obtain in practice). Another data-driven method that has emerged in recent years is OpInf, which uses time-domain state measurements (snapshots of the state variable), and then fits a particular nonlinear model (quadratic or quadratic-bilinear)
by computing the appropriate matrices.
%This is done using least-squares type methods.
Details of the operator
inference method can be found in  \cite{morPehW16} and also in more recent works such as \cite{morBenGKetal20}, \cite{morBenGHetal20}. 

In this contribution, we propose a data-driven procedure that can be used to infer quadratic-bilinear surrogate models from data. This can be viewed as an extension of the methods in \cite{morKarGA19a}  and in \cite{morGosKA20a}, which dealt with bilinear, and quadratic inference, separately. We propose a specific application for testing the method, i.e., electrical circuits with nonlinear components (such as diodes). 
%Although the dynamics characterizing the original model is strongly nonlinear, we perform lifting techniques to embed the original model into a quadratic-bilinear model.   
One aspect that distinguishes the OpInf framework from the extended Loewner-based frameworks that we propose in this contribution, is that the former requires
measurements of the whole state variable. Our framework requires only input-output measurements (transfer function measurements of higher-order generalized transfer functions). 
The paper is structured as follows; after the introduction, Section \ref{sec:ss_QB_Loew} introduces the class of QB systems with its generalized transfer functions. Here, we also briefly introduce the classical Loewner framework. Next, the newly-proposed method is introduced in Section \ref{sec:new_method}. Then, in Section \ref{sec:toy_circuit}, we first go through a simple example of a nonlinear circuit to show various reformulation of its structure. Section \ref{sec:rc_circuit} includes a detailed numerical study for applying the method to nonlinear ladder circuit, while Section \ref{sec:conc} gives the conclusions.

\section{Quadratic-bilinear systems and the Loewner framework}
\label{sec:ss_QB_Loew}

\subsection{State-space format and properties of QB systems}\label{sec:ss_QB}

We analyze in what follows dynamical systems as in (\ref{eq:nonlinsys}), with quadratic-bilinear (QB) nonlinearities for which $\bff$ is given as: $\bff(\bx(t),u(t))=\bQ(\bx(t)\otimes\bx(t))+\bN\bx(t)u(t)$. More precisely, let the state-space representation of such a system be given as:
%Consider the quadratic bilinear system of dimension $n$ 
\begin{equation}\label{eq:QBsys_ss}
\small
\left\{\begin{aligned}
\bE\dot{\bx}(t)&=\bA\bx(t)+\bQ(\bx(t)\otimes\bx(t))+\bN\bx(t)u(t)+\bB u(t),\\
y(t)&=\bC\bx(t),
\end{aligned}\right.,
\end{equation}
where $\bx(0)=\bx_0=\textbf{0}$ and the matrix $\bE\in\IR^{n\times n}$ is non-singular, $\bA\in\IR^{n\times n},~\bQ\in\IR^{n\times n^2},~\bN\in\IR^{n\times n},~\bB\in\IR^{n\times 1}$ and $\bC \in\IR^{1\times n}$. Moreover, assume that $\bQ$  satisfies the property $\bQ(\bv \otimes \bw) = \bQ(\bw \otimes \bv)$, i.e., it is represented in a "symmetrizable format".

%\subsection{Generalized transfer functions of QB systems}

%This is a procedure that gives the input to state generalized transfer function $\bG_{i}$ by denoting $\bPhi(s)=(s\bE-\bA)^{-1}\in\IC^{n\times n}$ as
%\begin{equation}
%\begin{aligned}
%G_{1}(s_1)&=\bPhi(s_1)\bB,\\
%G_{2}(s_{1},s_{2})&=\bPhi(s_1+s_2)\bQ(\bPhi(s_1)\bB\otimes\bPhi(s_2)\bB)\\
%&+ \bPhi(s_1+s_2)\bQ(\bPhi(s_2)\bB\otimes\bPhi(s_1)\bB) \\
%&+\frac{1}{2}\bPhi(s_1+s_2)\bN(\bPhi(s_1)\bB+\bPhi(s_2)\bB).
%\end{aligned}
%\end{equation}

%To obtain the input-output generalized frequency response functions, we multiply with $\bC$ 

The first two generalized symmetric transfer functions functions of a QB system as in (\ref{eq:QBsys_ss}) (sometimes referred to as Volterra kernels in the frequency domain) are:
\begin{equation}
\begin{aligned}
\small
H_{1}(s_1)&=\bC_{\ell}\bPhi(s_1)\bB,\\
H_{2}(s_{1},s_{2})&=\bC_{\ell}\bPhi(s_1+s_2)\bQ(\bPhi(s_1)\bB\otimes\bPhi(s_2)\bB)\\
&+\frac{1}{2}\bC\bPhi(s_1+s_2)\bN(\bPhi(s_1)\bB+\bPhi(s_2)\bB).
\end{aligned}
\end{equation}
For more details on deriving such functions we refer the reader to \cite{morBre13} and to \cite{GA18,morAntGH19}. An important property of these particular functions (sometimes called symmetric transfer functions), is that their samples can be inferred from the spectrum of the observed output when using a purely oscillatory control input.

% Hence, the second symmetric transfer function will have a simplified format (the first and second terms will collapse into one).

\subsection{The Loewner framework}\label{sec:loew}

We start with a brief summary of the Loenwer framework (LF) for fitting linear systems as in (\ref{eq:linsys}). For more details, we refer the reader to \cite{ALI17}. The starting point for LF is to collect measurements corresponding to the (first) transfer function, which can be inferred in practice from the first harmonic. The data are first partitioned into two disjoint subsets, as:
\begin{align}\label{data_Loew}
\begin{split}
{\textrm right \ data}&:(\lambda_j;w_j), ~j=1,\ldots,k,~{\textrm and}, \\
{\textrm left \ data}&:(\mu_i;v_i), ~i=1,\ldots,k,
\end{split}
\end{align} 
%(for simplicity all points are assumed distinct).
find the function
$\bH(s)$, such that the following interpolation conditions are (approximately) fulfilled:
\begin{equation} \label{interp_cond}
\bH(\mu_i)=v_i,~~~\bH(\lambda_j)=w_j.
\end{equation}
The Loewner matrix $\IL \in\IC^{k\times k}$ and the shifted Loewner matrix $\IL_s \in\IC^{k\times k}$ are defined as follows
\begin{equation} \label{Loew_mat}
\IL_{(i,j)}=\frac{v_i-w_j}{\mu_i-\lambda_j}, \ \IL_{s(i,j)}=
\frac{\mu_i v_i-\lambda_j w_j}{\mu_i-\lambda_j},
\end{equation}
while the data vectors $\IV, \IW^T \in \IR^k$ are introduced as
\begin{equation} \label{VW_vec}
\IV_{(i)}= v_i, \ \  \IW_{(j)} = w_j,~\text{for}~i,j=1,\ldots,k.
\end{equation}
The Loewner model is hence constructed as follows:
	\begin{align*}
	\bE=-\IL,~~ \bA=-\IL_s,~~ \bB=\IV,~~ \bC=\IW.
	\end{align*}
Provided that enough data is available, the pencil $(\IL_s,\,\IL)$ is often singular. In these cases, a singular value decomposition (SVD) of Loewner matrices is needed to compute projection matrices $\bX_r, \bY_r \in \IC^{k \times r}$. Here, $r<n$ represents the truncation index.

Then, the system matrices corresponding to a projected Loewner model of dimension $r$ can be computed using matrices $\bX_r$ and $\bY_r$, as: 
	\begin{equation}\label{Loew_red_lin}
	\small
	\hat{\bE} = -\bX_r^*\IL \bY_r, \ \  \hat{\bA} = -\bX_r^*\IL_s \bY_r, \ \
	\hat{\bB} = \bX_r^*\IV, \ \  \hat{\bC} = \IW \bY_r,
	\end{equation}
	and hence, directly find a state-space realization corresponding to the reduced-order system of equations
\begin{equation}\label{eq:linsys}
    \left\{\begin{aligned}
        \bhE \dot{\bhx}(t)&=\bhA\bx(t)+\bhB u(t),\ \ \hat{y}(t) =\bhC\bhx(t).
    \end{aligned}\right.
\end{equation}
More implementation details and properties on the LF procedure can be found in \cite{ALI17} and in \cite{morKarGA19a}.

\section{The proposed hybrid method based on the Loewner framework and LS solves}\label{sec:new_method}

%\subsection{The proposed method}

The idea is to recover all the operators from measurements. The Loewner framework is capable of recovering the linear part by providing a fitted realization of dimension $r$ as in (\ref{Loew_red_lin}), denoted with $(\bhE,\bhA,\bhB,\bhC)$. Now, based on these matrices, introduce $\bhPhi(s)=(s\bhE-\bhA)^{-1} \in \IC^{r \times r}$.

The sampling domain is denoted with $\Omega = \{(\zeta_1^{(i)},\zeta_2^{(i)}) \in \IC^2 \vert 1 \leq i \leq K\}$.
For a particular pair of sampling points $(\zeta_1^{(i)},\zeta_2^{(i)}) \in \IC^2$, we define the following quantities:
\begin{align}\label{eq:def_O_R}
\begin{split}
    \chO(\zeta_1^{(i)},\zeta_2^{(i)}) &:= \bhC\bhPhi(\zeta_1^{(i)}+\zeta_2^{(i)}) \in\IC^{1\times r}, \\
    \chR_q(\zeta_{1}^{(i)},\zeta_{2}^{(i)}) &:= (\bhPhi(\zeta_1^{(i)})\bhB\otimes\bhPhi(\zeta_2^{(i)})\bhB) \in\IC^{r^2 \times 1}, \\
    \chR_b(\zeta_1^{(i)},\zeta_2^{(i)}) &:= (\bhPhi(\zeta_1^{(i)})\bhB+\bhPhi(\zeta_2^{(i)})\bB) \in\IC^{r\times 1}.
    \end{split}
\end{align}
It is to be noted that the vectors introduced in (\ref{eq:def_O_R}) are computed solely in terms of the matrices $(\bhE,\bhA,\bhB,\bhC)$ corresponding to the data-driven Loewner surrogate model.
%The vectorization procedure\footnote{Row-wise vectorization: $vec(\textbf{P})=\left[\begin{array}{ccc}
%\textbf{P}(1,1:n) & \cdots & \textbf{P}(n,1:n)
%\end{array}\right]^T$} will allow to write the second kernel %$H_{2}(s_{1},s_{2})$ in a linear equation format with respect to the unknown nonlinear operators.
 Let $\bv \in \IC^K$ be the vector of data measurements, i.e., containing samples of the second symmetric transfer function $H_{2}(s_{1},s_{2})$ evaluated on the $\Omega$ grid. More precisely, let $\bv_i = H_{2}(\zeta_1^{(i)},\zeta_2^{(i)})$. Now, since we would like to fit a reduced-order QB model to interpolate the 2D data, it follows that $H_{2}(\zeta_1^{(i)},\zeta_2^{(i)}) = \hH_{2}(\zeta_1^{(i)},\zeta_2^{(i)})$. We can write:
\begin{align}
\vspace{-4mm}
&\underbrace{\hH_{2}(\zeta_1^{(i)},\zeta_2^{(i)})}_{\bv_i\in\IC}=\underbrace{\bhC\bhPhi(s_1+s_2)}_{\chO(\zeta_1^{(i)},\zeta_2^{(i)})\in\IC^{1\times r}}\bhQ\underbrace{(\bhPhi(\zeta_1^{(i)})\bhB\otimes\bhPhi(s_2)\bhB)}_{\chR_q(\zeta_1^{(i)},\zeta_2^{(i)})\in\IC^{r^2\times 1}}\\
&+(1/2)\underbrace{\bC\bPhi(\zeta_1^{(i)}+\zeta_2^{(i)})}_{\chO(\zeta_1^{(i)},\zeta_2^{(i)})\in\IC^{1\times r}}\bhN\underbrace{(\bhPhi(\zeta_1^{(i)})\bhB+\bhPhi(\zeta_2^{(i)})\bB)}_{\chR_b(\zeta_1^{(i)},\zeta_2^{(i)})\in\IC^{r\times 1}},
\end{align}
and hence it follows that:
\begin{align}
\bhH_{2}(\zeta_1^{(i)},\zeta_2^{(i)})&=\chO(\zeta_1^{(i)},\zeta_2^{(i)})\bhQ\chR_q(\zeta_1^{(i)},\zeta_2^{(i)})\\
&+(1/2)\chO(\zeta_1^{(i)},\zeta_2^{(i)})\bhN\chR_b(\zeta_1^{(i)},\zeta_2^{(i)})
\end{align}

\begin{definition}
Given a matrix $\bX \in \IC^{m \times n}$, we denote with $\text{vec}(\bX)$ the vector $(mn) \times 1$ computed as follows:
\begin{equation}
    \text{vec}(\bX) = \left[\begin{array}{ccc}
\bX(1,:) & \cdots & \bX(m,:)
\end{array}\right]^T \in \IC^{mn},
\end{equation}
where the MATLAB notation $\bX(k,:) \in \IC^{1 \times n}$ was used to refer to the kth row of $\bX$.
\end{definition}

The vectorization procedure adapted to the data-driven problem shown here is presented in (\ref{eq:vec_proc}). Let us denote with  $\cT \in \IC^{K \times (r^3+r^2)}$, the matrix for which the ith row is given by $\cT(i,:) = \left[ \begin{matrix} \cT_q(i,:) & \cT_b(i,:) \end{matrix} \right] \in \IC^{1 \times (r^3+r^2)}$
\begin{align}
\hspace{-2mm}& \begin{cases} \cT_q(i,:) = \left[\begin{array}{c}
\chO(\zeta_1^{(i)},\zeta_2^{(i)})\otimes\chR_q^T(\zeta_1^{(i)},\zeta_2^{(i)}) 
\end{array}\right] \in \IC^{K \times r^3}, \\
\cT_b(i,:) = \left[\begin{array}{c}
 \chO(\zeta_1^{(i)},\zeta_2^{(i)})\otimes\chR_b^T(\zeta_1^{(i)},\zeta_2^{(i)})
\end{array}\right] \in \IC^{K \times r^2}.
\end{cases} \hspace{-6mm}
\end{align}

\begin{table*}
 \vspace{-\baselineskip}
\begin{equation}\label{eq:vec_proc}
\normalsize
\begin{aligned}
\bv_i = \hH_{2}(\zeta_1^{(i)},\zeta_2^{(i)}) &=\chO(\zeta_1^{(i)},\zeta_2^{(i)})\bhQ\chR_q(\zeta_1^{(i)},\zeta_2^{(i)})+(1/2)\chO(\zeta_1^{(i)},\zeta_2^{(i)})\bhN\chR_b(\zeta_1^{(i)},\zeta_2^{(i)})\\
&=\left(\chO(\zeta_1^{(i)},\zeta_2^{(i)})\otimes\chR_q^T(\zeta_1^{(i)},\zeta_2^{(i)})\right)vec(\bhQ)+(1/2)\left(\chO(\zeta_1^{(i)},\zeta_2^{(i)})\otimes\chR_b^T(\zeta_1^{(i)},\zeta_2^{(i)})\right)vec(\bhN)\\
&=\underbrace{\left[\begin{array}{c|c}
\underbrace{\chO(\zeta_1^{(i)},\zeta_2^{(i)})\otimes\chR_q^T(\zeta_1^{(i)},\zeta_2^{(i)})}_{= \cT_q(i,:)} & \underbrace{\chO(\zeta_1^{(i)},\zeta_2^{(i)})\otimes\chR_b^T(\zeta_1^{(i)},\zeta_2^{(i)})}_{= \cT_b(i,:)}
\end{array}\right]}_{= \cT(i,:) \ \in \  \IC^{1 \times (r^3+r^2)}} \cdot
\underbrace{\left[\begin{array}{c}
\text{vec}(\bhQ)\\\hline
(1/2)\text{vec}(\bhN)
\end{array}\right]}_{= \bz \ \in \ \IC^{(r^3+r^2) \times 1}}
\end{aligned}
\vspace{-3mm}
\end{equation} 
 \vspace{0.5\baselineskip}
  \hrule
\end{table*}
Now, from (\ref{eq:vec_proc}), by varying the index $i$ such as $1 \leq i \leq K$, it follows that we can put together a linear system of equations in $r^3+r^2$ unknowns as follows:
\begin{equation}\label{eq:LinSysEqs}
    \cT \bz  = \bv,
\end{equation}
where  $\bz = \left[\begin{matrix}
\text{vec}(\bhQ)^T & 
(1/2)\text{vec}(\bhN)^T
\end{matrix}\right]^T$ is the vector of variables which contains the entries of the vectorized operators of the surrogate reduced-order QB system. In order to ensure an over-determined linear system of equations, we clearly need to have enough data measurements, i.e., the condition $K\geq r^3+r^2$ needs to hold true. Then, we can employ a direct solution of system (\ref{eq:LinSysEqs}), e.g., by means of the Moore-Penrose pseudo-inverse or by using Gaussian elimination. If the matrix $\cT$ isindeed of full column rank, then it does not matter what procedure is chosen (for the direct solve). However, in most cases, the matrix $\cT$ is not of full column rank, and hence direct solves need to be carefully dealt with (by introducing regularization techniques). In what follows, we will use a truncated singular value decomposition (tSVD) approach. This is an attractive and powerful method since it uses the optimal rank-k approximation of the SVD (in the 2 norm). Such approach has been already used for applying OpInf \cite{morBenGHetal20}, together with the Tikhonov regularization scheme \cite{morPehW16} and the tQR (truncated QR decomposition) approach.

\begin{remark}
Finding a viable solution for the least-squares problem stated in (\ref{eq:LinSysEqs}) is not a straightforward task and can be computationally challenging task because of the ill-conditioning of matrix $\cT$ (the column rank of matrix $\cT$ is sometimes much smaller than $r^3+r^2$). Another challenge is the computational cost which grows cubically in the order $r$ of the reduced-order system.
\end{remark}

%To correctly estimate the operators of such a surrogate reduced-order QB system, we need at least $K\geq n^3+n^2$ measurements, and the above system can be expanded as

%\begin{remark}(Under-determined)
%It turns out that $rank(M)<n^3+n^2$, and this deficient problem provides only under-determined solutions of the least squares problem that cannot be led to the complete identification of the nonlinear operators.
%\end{remark}

\section{A simple model of a non-linear circuit}
\label{sec:toy_circuit}

We consider a simple circuit constructed from two blocks connected in series. Each block contains a capacitor in parallel with a diode. This is modeled as a dynamical system in two variables given by the voltage drops on each block. The input is given by the current through the circuit, while the observed output is the sum of the two variables. More precisely, we can write the differential equations characterizing the dynamics, as follows:
\begin{equation}\label{eq:nonlin_toy_sys1}
\left\{\begin{aligned}
C_{1}\frac{dV_{1}(t)}{dt}&=I(t)-I_{r_{1}}\left(e^{\frac{1}{V_{t_{1}}}V_{1}(t)}-1\right),\\
C_{2}\frac{dV_{2}(t)}{dt}&=I(t)-I_{r_{2}}\left(e^{\frac{1}{V_{t_{2}}}V_{2}(t)}-1\right),
\end{aligned}\right.
\end{equation}
while the output is $y(t)=V_{1}(t)+V_{2}(t)$. The capacitance  values are denoted with $C_{i}, 1 \leq i \leq 2$, while other constants are denoted with $I_{r_{i}}$ and $V_{t_{i}}$. Next, we introduce $x_{1}(t)=V_{1}(t)/V_{t_{1}},~x_{2}(t)=V_{2}(t)/V_{t_{2}}$, and let $a=\frac{1}{C_{1}V_{t_{1}}},~b=\frac{1}{C_{2}V_{t_{2}}},~c=I_{r_{1}},~d=I_{r_{2}}$. Using all of these, the nonlinear system in (\ref{eq:nonlin_toy_sys1}) is rewritten as:
\begin{equation}\label{eq:nonlin_toy_sys2}
\left\{\begin{aligned}
\dot{x}_{1}(t)&=aI(t)-ac\left(e^{x_{1}(t)}-1\right),\\
\dot{x}_{2}(t)&=bI(t)-bd\left(e^{x_{2}(t)}-1\right),\\
y(t)&=x_{1}(t)V_{t_{1}}+x_{2}(t)V_{t_{2}}.
\end{aligned}\right.
\end{equation}

\subsection{Analysis based on Taylor series truncation}

%The original nonlinear system is described by the equations above.

%From (\ref{eq:nonlin_toy_sys2}), we extract the linear counterpart of the nonlinear system, with the realization given by:
%\begin{align*}
%    \bA_N = \left[ \begin{matrix}
%    0 & 0 \\ 0 & 0
%    \end{matrix} \right],\ \     \bB_N = \left[ \begin{matrix}
%    a \\ b
%    \end{matrix} \right], \ \ \     \bC_N = \left[ \begin{matrix}
%    V_{t_{1}} & V_{t_{2}}
%    \end{matrix} \right].
%\end{align*}

%\begin{equation}
%    e^{x_i(t)} = 1 + x_i(t)+0.5x_i^2(t)+\cdots
%\end{equation}

%\subsubsection{Linearization by Taylor series truncation}

The methods discussed here are inexact, i.e., based on approximation (e.g., on truncating the Taylor series associated to the nonlinearity). First, by using a truncated Taylor series given by the formula $e^{x_i(t)} \approx 1 + x_i(t)$, and substitute it in (\ref{eq:nonlin_toy_sys2}), it follows that the following linear dynamical system is obtained:
\begin{equation}
\left\{\begin{aligned}
\dot{x}_{1}(t)&=aI(t)-ac x_1(t),\\
\dot{x}_{2}(t)&=bI(t)-bd x_2(t),\\
y(t)&=x_{1}(t)V_{t_{1}}+x_{2}(t)V_{t_{2}}+R\cdot I(t).
\end{aligned}\right.
\end{equation}
Then, identify the following linear realization:
\begin{align*}
    \bA_L = \left[ \begin{matrix}
    -ac & 0 \\ 0 & - bd
    \end{matrix} \right],\ \     \bB_L = \left[ \begin{matrix}
    a \\ b
    \end{matrix} \right], \ \ \     \bC_L = \left[ \begin{matrix}
    V_{t_{1}} & V_{t_{2}}
    \end{matrix} \right].
\end{align*}

%In this way, we recover the linear part corresponding to the linearized nonlinear system...and not the linear part of the lifted model! The former one gives us the first linear TF.

%\subsubsection{Bilinearization by Carleman's method}

Next, we discuss Carleman's linearization proposed in  \cite{Car32}. It is a method used to embed a nonlinear system of differential equations of finite dimension
into a system of bilinear differential equations of infinite dimension. By truncating the hence obtained bilinear system at finite orders, one finds a systematic way of achieving an arbitrary-order approximation. By following this procedure, introduce the augmented new state by stopping at quadratic terms (and ignoring all other higher powers), as
\begin{equation}
    \bx^C = \left[ \begin{matrix}\bx \\ \bx \otimes \bx \end{matrix} \right] = \left[ \begin{matrix} x_1 & x_2 &  x_1^2 &  x_1 x_2 &  x_2 x_1 &  x_2^2 \end{matrix} \right]^T,
\end{equation}
and appropriately compute the derivatives of the last four entries as, for example, of:
\begin{align}
\begin{split}
    \dot{x}^C_3(t) &= \frac{d}{dt} x_1^2(t) = 2 x_1(t) \dot{x}_1(t) \\ %&= 2 x_1(t)  \left(aI(t)-ac x_1(t)-0.5ac x_1^2(t) \right) \\
    &=  2 a \bx_1(t)  I(t)-2 ac x_1^2(t)-ac x_1^3(t).
    \end{split}
\end{align}
Finally, the approximation steps follows, i.e., we neglect the powers in $x_i(t)$ higher than 3, and write:
\begin{align}
\begin{split}
    \dot{x}^C_3(t) &\approx  2 a x_1(t)  I(t)-2 ac x_1^2(t) \\
    &\Rightarrow   \dot{x}^C_3(t) \approx  2 a x^C_1(t)  I(t)-2 ac x_2^C(t),
    \end{split}
\end{align}
which includes a linear term, i.e., $ x_2^C(t)$, and a bilinear term: $x^C_1(t)  I(t)$. The same procedure is applied for the other entries of the derivative of the new state vector  $\dot{\bx}^C(t)$. Hence, an approximate bilinear systems is derived:
\begin{equation*}
\left\{\begin{aligned}
 \dot{x}^C(t)&=\bA\bx^C(t)+\bN\bx^C(t)u(t)+\bB u(t),\ \ y(t)=\bC\bx^C(t).
\end{aligned}\right.
\end{equation*}

\subsection{Polynomial lifting}\label{sec:PolFit}

The first step towards implementing a "lifting approach" is to identify the quantities that depend non-linearly on the original states, and to consider them as new artificial states. For example, we  introduce the auxiliary variables 
\begin{equation}
x_{3}(t):=e^{x_{1}(t)}-1, \ \ \ \ x_{4}(t):=e^{x_{2}(t)}-1.
\end{equation}
In this way, we also enforce zero initial conditions. The augmented system is hence written as:
\begin{equation}\label{eq:lift_sys}
%\small
\hspace{2mm} \left\{\begin{aligned}
\dot{x}_{1}(t)&=aI(t)-acx_{3}(t),\ \ \dot{x}_{2}(t)=bI(t)-bdx_{4}(t),\\
\dot{x}_{3}(t)&=-acx_{3}(t)-acx_{3}^2(t)+aI(t)+ax_{3}(t)I(t)\\
\dot{x}_{4}(t)&=-bdx_{4}(t)-bdx_{4}^2(t)+bI(t)+bx_{4}(t)I(t)\\
y(t)&=x_{1}(t)V_{t_{1}}+x_{2}(t)V_{t_{2}}+R\cdot I(t).
\end{aligned}\right.
\end{equation}
The system in (\ref{eq:lift_sys}) can be written in the general QB form:
\begin{equation*}
\left\{\begin{aligned}
\dot{\bx}(t)&=\bA\bx(t)+\bQ(\bx(t)\otimes\bx(t))+\bN\bx(t)I(t)+\bB I(t),\\
y(t)&=\bC\bx(t),
\end{aligned}\right.\\
\end{equation*}
where the matrices are defined next:
\tiny
\begin{equation*}
\begin{aligned}
\bA&=\left[\begin{array}{cccc}
0 & 0& -ac& 0\\
0 & 0& 0& -bd\\
0 & 0& -ac& 0\\
0 & 0& 0& -bd
\end{array}\right],
\bN=\left[\begin{array}{cccc}
0 & 0& 0& 0\\
0 & 0& 0& 0\\
0 & 0& a& 0\\
0 & 0& 0& b
\end{array}\right],~\bB=\left[\begin{array}{c}
a\\
b\\
a\\
b
\end{array}\right], \\
\bC&=\left[\begin{array}{cccc}
V_{t_{1}} \\ V_{t_{2}} \\ 0 \\ 0\end{array}\right]^T, \bQ=\left[\begin{array}{c|cccc}
\textbf{0}_{2\times 12} & \textbf{0}_{2\times 4}\\\hline
\textbf{0}_{2\times 12} & \bQ_{22}\\
\end{array}\right],~\bQ_{22}=\left[\begin{array}{cccc}
-ac & 0 & 0 & 0\\
0 & 0 & 0 & -bd\end{array}\right]
\end{aligned}
\end{equation*}
\normalsize
The first two symmetric transfer functions, where $\bPhi(s)=(s\bE-\bA)^{-1}\in\mathbb{C}^{n\times n}$, can be written as:
\small
\begin{align}
\begin{split}
H_{1}(s)&=\bC\bPhi(s)\bB=\frac{V_{t_{1}}}{C_1V_{t_{1}}s+I_{r_{1}}}+\frac{V_{t_{2}}}{C_2V_{t_{2}}s+I_{r_{2}}},  \\
H_{2}(s,s)&=\bC\bPhi(2s)\bQ\left[\bPhi(s)\bB\otimes\bPhi(s)\bB\right]+\bC\bPhi(2s)\bN\bPhi(s)\bB \nonumber \\
     &= - \frac{{I}_{r_1} {V_{t_1}}}{2\,{\left({I}_{r_1}+C_{1}\,{V_{t_1}}\,s\right)}^2\,\left({I}_{r_1}+2\,C_{1}\,V_{t_1}\,s\right)}\\
     &-\frac{{I}_{r_2} V_{t_2}}{2\,\left({I}_{r_2}+C_{2}\,V_{t_2}\,s\right)^2\,\left({I}_{r_2}+2\,C_{2}\,V_{t_2}\,s\right)} 
 \end{split}
\end{align}
\normalsize

% {I}{r_1} should be {I}_{r_1} ...aaa okokoko
% same for Vt stuff etc.

%{\color{blue} The mathrm format above is not necessary and it is different than $H_1(s)$ format...so pls change}

The parameters can be recovered from LF by considering the 2nd kernel as a univariate rational function, but the pole residue form needs a special treatment and is usually quite challenging. Similar studies have been proposed in \cite{morKGA2019bil,morGosKA20a} for inferring bilinear or quadratic systems respectively, where an improvement towards bilinear identification was shown in \cite{morKarGA21a} and in the current study. 

% there is not much space here...before Section 5
% fixed 
%\vspace{-2mm}

\section{Numerical experiments: a nonlinear RC Ladder circuit}
\label{sec:rc_circuit}

We analyze a nonlinear RC-ladder electronic circuit first introduced in \cite{Chen99}. Various variants of this model were also mentioned in other MOR works,i.e., \cite{Gu11} and \cite{morBre13}. This nonlinear first-order system models a resistor-capacitor network that exhibits a nonlinear behaviour caused by the nonlinear resistors consisting of a parallel connected resistor with a diode.
%or the nonlinear resistors connected parallel to the capacitor.

%First, we discuss the modeling of an RC circuit, where the nonlinear resistors consist of a parallel connected resistor with a diode as shown in the above figure.

As presented in \cite{Chen99}, the underlying model is given by a SISO system of the form:
\small
\begin{align}\label{eq:RCladder}
\dot{x}(t) = \left[ \begin{matrix} -g(x_1(t)) - g(x_1(t) - x_2(t)) \\ g(x_1(t)-x_2(t)) - g(x_2(t)-x_3(t)) \\ \vdots \\ g(x_{k-1}(t) - x_k(t)) - g(x_k(t) - x_{x+1}(t)) \\ \vdots \\ g(x_{N-1}(t) - x_N(t)) \end{matrix} \right] +\left[ \begin{matrix}  u(t) \\ 0 \\ \vdots \\ 0 \\ \vdots \\ 0  \end{matrix} \right],
\end{align}
\normalsize
with $y(t) = x_1(t)$, where the mapping $g$ is given by $g:\mathbb{R} \to \mathbb{R}$ defined as $g(x_i) = g_D(x_i) + x_i$, which combines the effect of a diode and a resistor. The non-linearity $g_D$ models a diode as a nonlinear resistor, based on the classical Shockley model:
\begin{equation}\label{eq:Shockley}
g_D(x_i) = i_S (\exp(u_P x_i) - 1),
\end{equation}
with material parameters $i_S > 0$ and $u_P > 0$. For this benchmark, the parameters are selected as follows: $i_S$ = 1 and $u_P = 40$ as in \cite{Chen99}. By substituting these values into (\ref{eq:Shockley}), we get that  $g_D(x_i) = \exp(40 x_i) - 1$, and hence it follows that $g(x_i) = \exp(40 x_i) +x_i - 1$.
%By following the work of Y. Chen, clearly the nonlinear mapping above can be approximated by Taylor series expansion (around $0$) and keeping only the first terms:
%\begin{enumerate}
%	\item Linear approximation: $g_1(x_i) = 41 x_i$.
%	\item Quadratic approximation: $g_2(x_i) = 41 x_i+800 x_i^2$.
%\end{enumerate}
In what follows, we will apply the proposed methods for:

%\subsubsection{Linear approximation}

%By completely eliminating the higher-order terms from the Taylor expansion, we are using here the following linear approximation to $g_D$, i.e. $g_D(x_i) = 40x_i$, and hence $g(x_i) = g_1(x_i) = 41 x_i$.

%{\color{blue} We do not use this, or?! Then we can just remove the above paragraph.}
%\textcolor{red}{Yes, I think we can remove this part!}

%\begin{align}
%\dot{x}(t) = \left[ \begin{matrix} -82 x_1(t) + 41 x_2(t) \\  41 x_1(t)-82 x_2(t) + 41 x_3(t)  \\ \vdots \\ 41 x_{k-1}(t)-82 x_k(t) + 41 x_{k+1}(t) \\ \vdots \\ 41x_{N-1}(t) - 41 x_N(t) \end{matrix} \right] +\left[ \begin{matrix}  u(t) \\ 0 \\ \vdots \\ 0 \\ \vdots \\ 0  \end{matrix} \right],
%\end{align}

%\subsubsection{Quadratic approximation}

%By  eliminating the higher-order terms starting with the cubic ones, from the Taylor expansion, we are using here the following linear approximation to $g_D$, i.e. $g_D(x_i) = 40x_i+800x_i^2$, and hence 
%\begin{equation}
%g(x_i) = g_1(x_i) = 41 x_i+800x_i^2.
%\end{equation}

\subsection{Bilinear treatment via Carleman's approach}

The original nonlinear system is transformed into a bilinear system by means of Carleman linearization, as originally shown in \cite{Chen99} and later in \cite{morBre13}. The matrices are:
\begin{equation}\label{eq:bilearRCladder}
\small
    \begin{aligned}
     &\bA=\left[\begin{array}{cc}
        \bA_1  & \frac{1}{2}\bA_2 \\
         \textbf{0} & \bA_1\otimes\bI+\bI\otimes\bA_1\end{array}\right],~\bx=\left[\begin{array}{c}
              \bv  \\
              \bv\otimes\bv\end{array}\right],\\
              &\bN=\left[\begin{array}{cc}
        \textbf{0}  & \textbf{0} \\
         \bB\otimes\bI+\bI\otimes\bB & \textbf{0}\end{array}\right],~\bB=\left[\begin{array}{c}
        \bb\\\textbf{0}
        \end{array}\right],~\bC=\left[\begin{array}{c}
             \bc  \\
             \textbf{0}
        \end{array}\right]^T.
    \end{aligned}
\end{equation}
Consequently, the resulting bilinear system has dimension $n^2+n$ with $n$ as the number of circuit blocks of the original system. More details on the structure of the involved matrices in Eq.\;\eqref{eq:bilearRCladder} can be found in \cite{morBre13}.
%Originally, let $N = 10$ and, hence, the corresponding bilinear system is of dimension $N_B = N^2+N = 110$.

%Choose the control input to be $u(t) =  5 \cos( \pi/5 t)+5$.

\subsection{Lifting to Quadratic-Bilinear form}

Analogues with the example in Sec.\;\ref{sec:PolFit}, the original RC-ladder model can be lifted to an equivalent quadratic bilinear model. The introduced additional state variables $\bx_1=\upsilon_1$ and $x_i=\upsilon_i-\upsilon_{i+1}$ followed by introducing the additional state variables  $z_1=e^{-40\upsilon_1}-1$ and $z_i=e^{40x_i}$ can transform equivalently the original system Eq.\;\eqref{eq:RCladder} to a quadratic-bilinear form Eq.\;\eqref{eq:QBsys_ss} with dimension $2n$ \cite{morBre13}.

\begin{figure}[h!]
    \centering
    \includegraphics[scale=0.27]{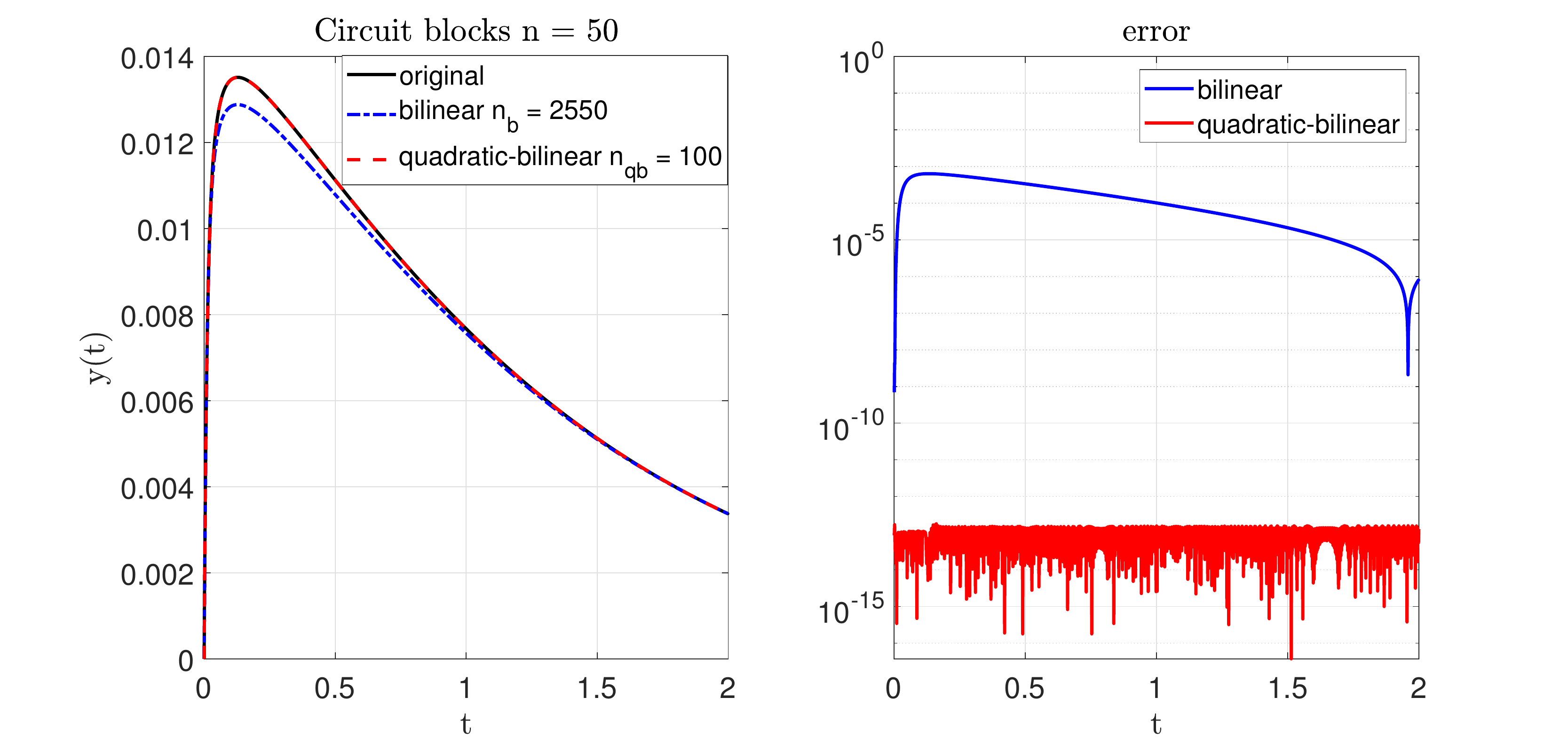}
        \vspace{-5mm}
    \caption{\small The original RC-ladder model with $n=50$ circuit blocks. The lifted QB model is equivalent with the original nonlinear. The bilinear model of dimension $n^2+n$ offers good approximation only for relative small input amplitudes. \normalsize}
    \label{fig:equivalence}
        \vspace{-2mm}
\end{figure}

In Fig.\;\ref{fig:equivalence} the original nonlinear system along with the equivalent quadratic-bilinear and the approximated bilinear are depicted. The numerical difference between the original and the quadratic-bilinear has reached machine precision where the bilinear for this amplitude starts to differ significantly.

The aim is to achieve MOR by measuring the first symmetric Volterra kernels from input-output time-domain simulations. As we infer the operators from the $2$nd kernel a double tone input is considered. The scheme for kernel separation and harmonic indexing remains the same as in \cite{morKarGA21a}. By simulating the original nonlinear model in the time domain under the excitation of a double-tone harmonic input, an accurate separation of kernels can be achieved (the Fourier transform is indeed accurate). Here, as we want to illustrate the efficacy of the proposed method by inferring the operators from the $2$nd Kernel, we assume a perfect measurement setup. 

\begin{figure}[h!]
    \centering
    \includegraphics[scale=0.27]{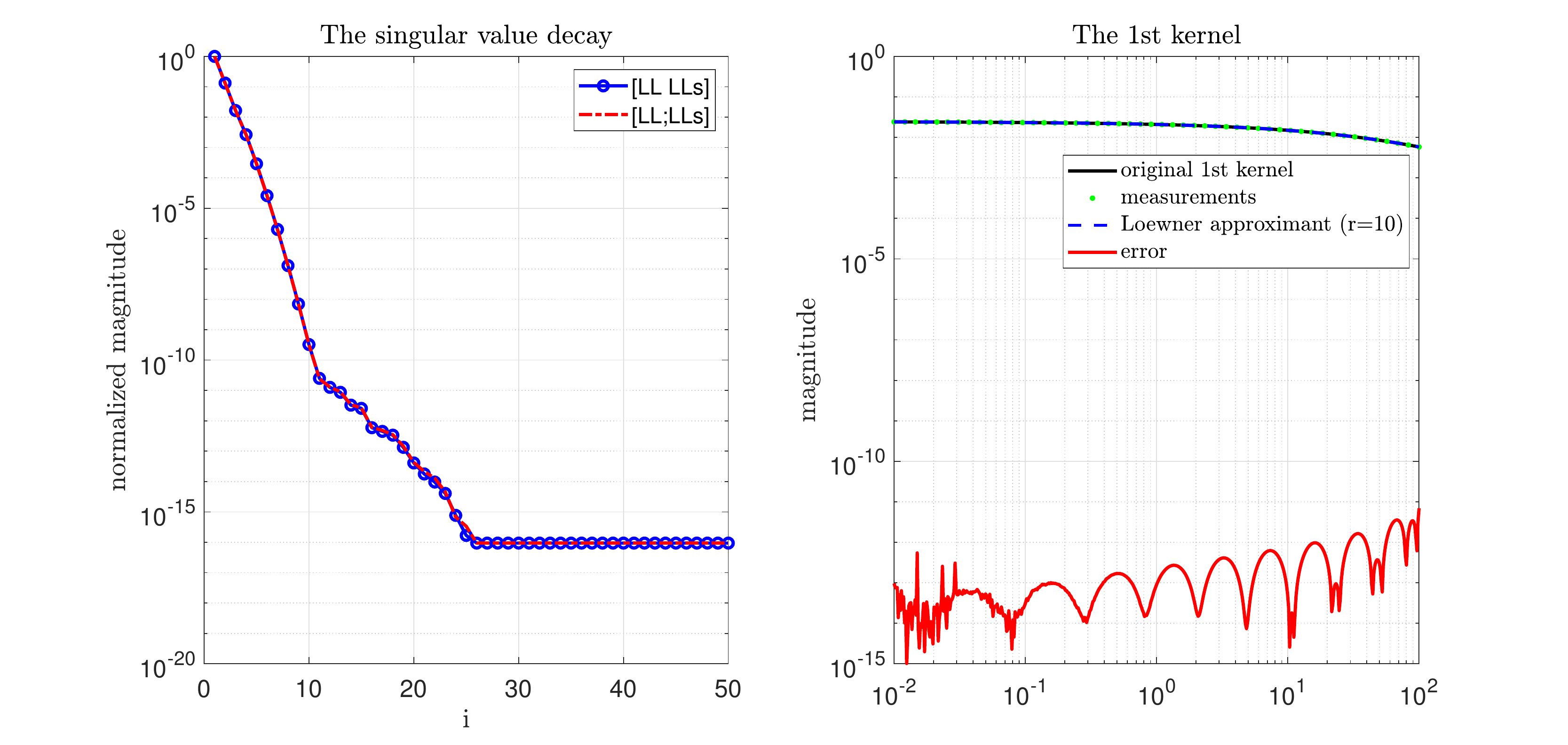}
    \vspace{-5mm}
    \caption{\small Left pane: The Loewner singular value decay (used to decide on the reduced order). Right pane: The reduced-order Loewner interpolant computed from the $1$st kernel that can reach machine precision approximation. \normalsize}
    \label{fig:fig2}
        \vspace{-2mm}
\end{figure}

\textbf{Step 1:} The first harmonic can be measured under excitation with a single tone input. Then, measurements of the first kernel $H_{1}(s_1)$ (e.g., the magnitude and phase) can be derived. The Loewner framework constructs a low-order rational interpolant and identifies the minimal linear sub-system of order $r$. In Fig.\;\ref{fig:fig2} and on the left pane, the Loewner singular value decay offers the criterion for reduction. The order $r=10$ was chosen, as the $11$th singular value is close to machine precision. Hence, it is an indicator for recovering the original linear dynamics. Denote with $\Sigma_{\texttt{lin}}: (\hat{\bA},\hat{\bB},\hat{\bC})$ the linear reduced system of order $r=10$.

In Fig.\;\ref{fig:fig2} and on the right pane, the approximation results are depicted.

\textbf{Step 2:} By having access to the reduced operators of the linear system $\hat{\Sigma}_\texttt{lin}$, we can infer the remaining nonlinear quadratic and bilinear operators from Eq.\;\eqref{eq:vec_proc}. The second harmonic can be measured with a double tone excitation. Thus, measurements of the second kernel $H_{2}(s_1,s_2)$ over the whole complex domain of definition can be collected repetitively. It is important here to mention that the amount of measurements is related to the reduced dimension $r$, where at least $K\geq(r^3+r^2)$ measurements ensures enough data, for the solution of Eq.\;\eqref{eq:LinSysEqs}. By enforcing quadratic symmetries (the matrix $\hat{\bQ}$ is set to satisfy the property $\hat{\bQ}(\bw\otimes\bv)=\hat{\bQ}(\bv\otimes\bw)$), the complexity can be further reduced. Using some algebraic adjustments, another simplification can be performed by replacing the symmetric Kronecker product with the asymmetric one ($\otimes'$ as in \cite{morBenGKetal20}). Solving for the vector $\bz$ yields a reduced-order QB system $\hat{\Sigma}_{QB}=(\hat{\bA},\hat{\bQ},\hat{\bN},\hat{\bB},\hat{\bC})$. 
\begin{figure}[h!]
    \centering
    \includegraphics[scale=0.27]{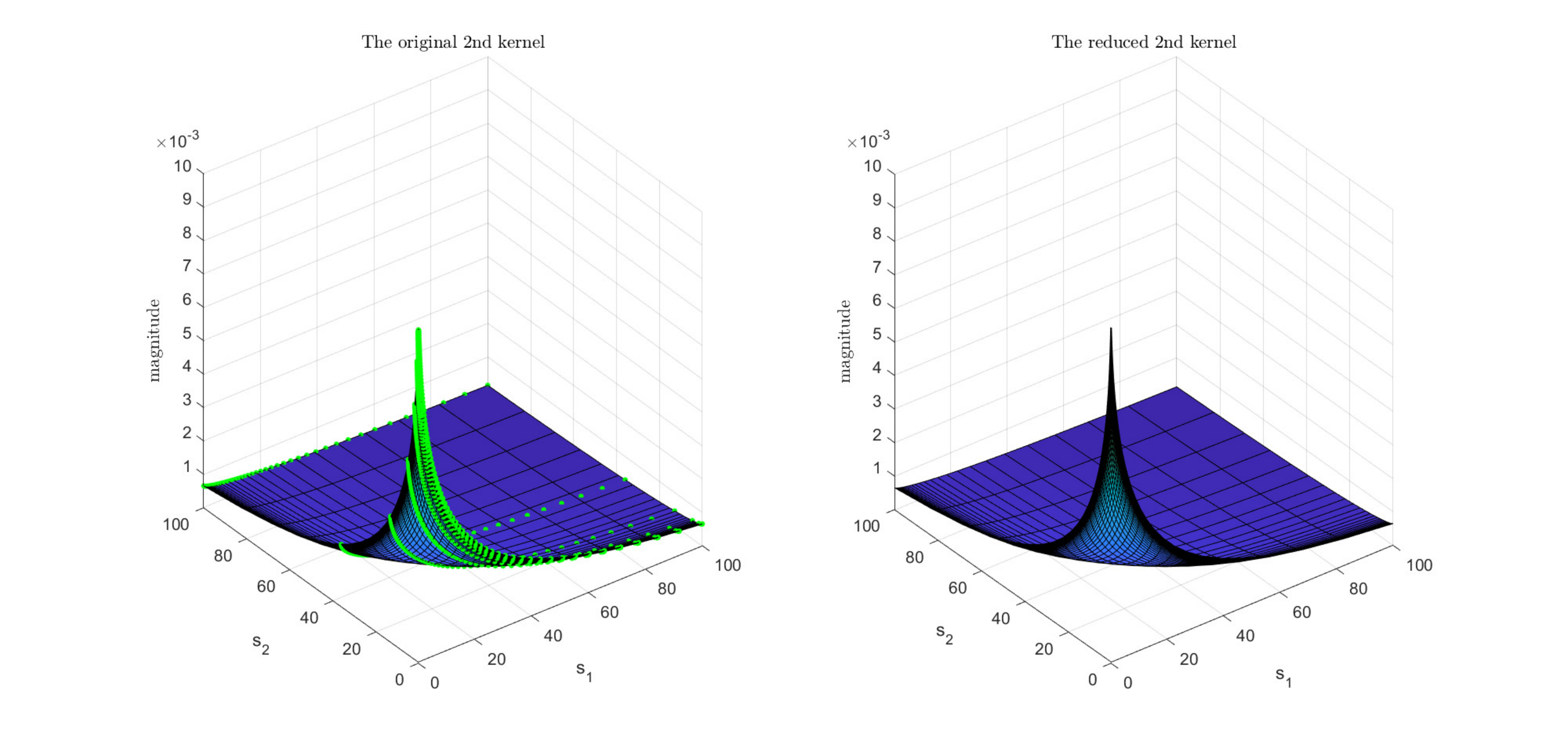}
        \vspace{-5mm}
    \caption{\small Left pane: The original 2nd kernel along with the measurements (green dots). Right pane: The 2nd kernel of the reduced system. $\Vert H_{2}(s_1,s_2)-\hat{H}_{2}(s_1,s_2) \Vert_{\infty}\sim 10^{-8}$.\normalsize }
    \label{fig:fig3}
        \vspace{-2mm}
\end{figure}
Finally, in Fig.\;\ref{fig:fig4} the time domain solution is depicted and is compared to the original nonlinear response.  
\begin{figure}[h!]
    \centering
    \includegraphics[scale=0.27]{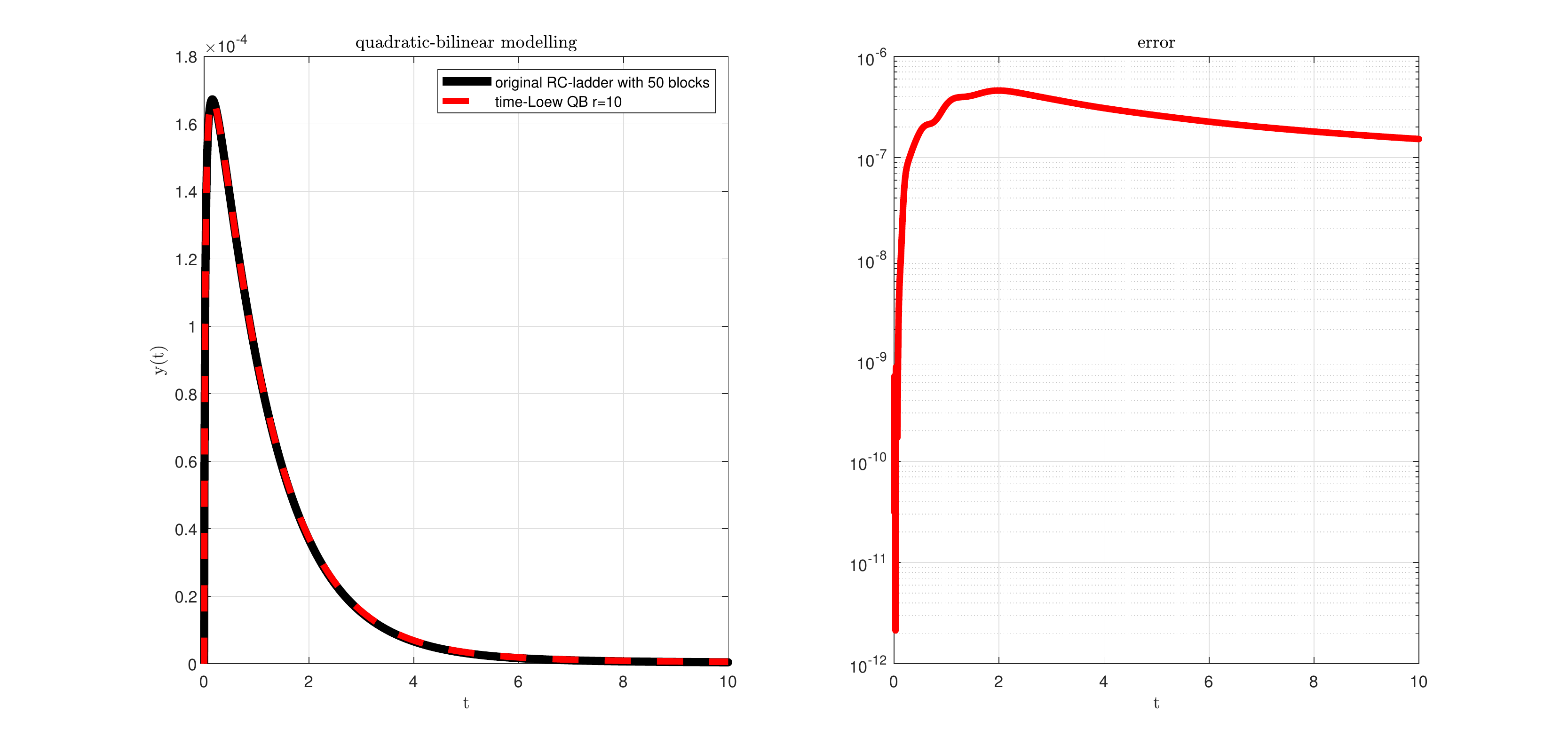}
        \vspace{-5mm}
    \caption{\small Left pane: The original nonlinear RC-ladder along with the approximant that is constructed with the new LoewQB method. Right pane: The absolute error over a dense grid. \normalsize}
    \label{fig:fig4}
    \vspace{-2mm}
\end{figure}

The proposed method successfully constructs a reduced quadratic-bilinear model from input-output time domain data that accurately approximates the response of the original nonlinear system. The simulations are performed with the multi-step Runge Kutta scheme (e.g., using \texttt{ODE45} in Matlab) with input $u(t)=0.01e^{-t}$, and the maximum error is $\Vert (y(t)-y_{r}(t))\Vert_{\infty}\sim 10^{-6}$. 
\section{Conclusion and future endeavours}
\label{sec:conc}

A non-intrusive data-driven method that constructs nonlinear models with the quadratic-bilinear structure from input-output time-domain data was presented. The new method was based on the Loewner and Volterra frameworks. The second symmetric kernel can be inferred as the $2$nd harmonic of the transformed output and hence, the problem of estimating the nonlinear operators can be resolved into a linear LS system. Although this LS system may be under-determined, it contains enough information for estimating the nonlinear operators with the Moore-Penrose pseudo-inverse (truncated SVD), thus providing good approximations. The use of higher harmonics (e.g., $3$rd kernel, etc.) that results in a nonlinear optimization problem along with the use of other regularization techniques as the one proposed by Tikhonov, will be the topics of future research endeavors in connection also with the analysis on the sensitivity of noise that real data contain. 
% looks good to me! no further comments

%%%%%%%%%%%%%%%%%%%%%%%%%%%%%%%%%%%%%%%%%%%%%%%%%%%%%%%%%%%%%%%%%%%%%%%%%%%%%%%%
% *** REFERENCES ***                                                           %
%%%%%%%%%%%%%%%%%%%%%%%%%%%%%%%%%%%%%%%%%%%%%%%%%%%%%%%%%%%%%%%%%%%%%%%%%%%%%%%%

%\addcontentsline{toc}{section}{References}
%\bibliographystyle{plainurl}

%\footnotesize
%\bibliographystyle{ifacconf}
\small
\bibliographystyle{plain}
\bibliography{tLoewQBKGA}
\end{document}